\documentclass[9pt,twocolumn,twoside]{osajnl}
\pdfoutput=1

\journal{ol} 

\setboolean{shortarticle}{true}

\usepackage{amsmath}
\usepackage{amsfonts}
\usepackage{bbold}
\usepackage{graphicx}

\usepackage{pdfpages}

\newcommand{\mhz}{\;\mathrm{MHz}}
\newcommand{\tdsh}{\text{-}}
\newcommand{\envb}[1]{\bar{b}_{j,#1}}

\title{Nonlocal magnon entanglement generation in coupled hybrid cavity systems}

\author[1]{Da-Wei Luo}
\author[1,*]{Xiao-Feng Qian}
\author[1,$\dagger$]{Ting Yu}

\affil[1]{
Department of Physics and Center for Quantum Science and Engineering, Stevens Institute of Technology, Hoboken, New Jersey 07030, USA
}

\affil[*]{Corresponding Author: xiaofeng.qian@stevens.edu}
\affil[$\dagger$]{Ting.Yu@stevens.edu}

\begin{abstract}
We investigate dynamical generation of macroscopic nonlocal entanglements between two remote massive magnon-superconducting-circuit hybrid systems. Two fiber-coupled microwave cavities are employed to serve as an interaction channel connecting two sets of macroscopic hybrid units each containing a magnon (hosted by a Yttrium-Iron-Garnet sphere) and a superconducting-circuit qubit. Surprisingly, it is found that stronger coupling does not necessarily mean faster entanglement generation. The proposed hybrid system allows the existence of an optimal fiber coupling strength that requests the shortest amount of time to generate a systematic maximal entanglement. Our theoretical results are shown to be within the scope of specific parameters that can be achieved with current technology. The noise effects on the implementation of systems are also treated in a general environment suggesting the robustness of entanglement generation. Our discrete-variable qubit-like entanglement theory of magnons may lead to direct applications in various quantum information tasks.
\end{abstract}
\setboolean{displaycopyright}{true}

\begin{document}

\maketitle

\section{Introduction}
Nonlocal entanglement between remote quantum objects is of crucial importance to various quantum information schemes, especially in the proposals of quantum communication, quantum cryptography, quantum sensing, etc.~\cite{Nielsen2000a}. Unfortunately, quantum entanglement is typically fragile due to the notorious effect of environment-induced disentanglement ~\cite{Yu2009a}. Therefore, the search for optimal physical systems that permit robust entanglement has never ceased. Macroscopic systems are believed to be one of the promising candidates. This has triggered various investigations of entanglement in massive quantum systems including quantum spin-state of Caesium gas samples~\cite{Julsgaard01a}, quantum nano optical-mechanical cavities~\cite{Riedinger2018a,Ockeloen-Korppi2018a}, and even in classical optical systems~\cite{Qian2017a}. Recently, collective spin excitations, termed as ``magnons", in macroscopic ferromagnetic materials such as Yttrium-Iron-Garnet (YIG) spheres shed a new light on developing a robust macroscopic quantum system that possesses certain favorable features for quantum information science and technology \cite{Tabuchi2014a, Tabuchi2016a, Osada2016a, Haigh2016a, Zhang2016a, Li2018a, Tabuchi2015a, Lachance-Quirion2017a, you_nat1, you_nat2}. Two notable attributes of the magnon systems are their long lifetimes and great tunability \cite{Zhang2016a}. In addition, these systems are demonstrated to be able to coherently exchange quantum information with other major types of qubit-candidate systems including microwave photons in the strong and even ultra-strong coupling regime through a cavity \cite{Tabuchi2014a, Soykal2010a, Imamoglu2009a}, superconducting circuits in the microwave regime \cite{Tabuchi2016a, Tabuchi2015a, Lachance-Quirion2017a}, optical photons through magneto-optical interactions ~\cite{Osada2016a, Haigh2016a}, etc. They can also induce Kerr nonlinear effects and display bi-stability~\cite{Wang2018a} for quantum operation. 

\begin{figure}
\centering
    \includegraphics[scale=.45]{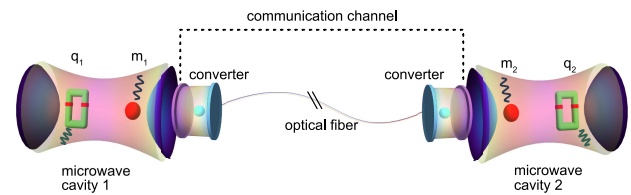}
    \caption{(Color online) Schematic representation of the system under consideration, consisting of two coupled microwave cavities. The communication channel between the two microwave cavities is realized by two reversible microwave-optical photon converters and an optical fiber. Inside each cavity is a YIG ferromagnetic sphere ($m_{1,2}$) and a SQ ($q_{1,2}$), both coupled to the cavity model.}\label{fig_schm}
\end{figure}

Due to the magnon's oscillator nature, continuous-variable entanglement has been explored in different cases, e.g., generations of two-magnon entanglement have been proposed both in a single cavity~\cite{Li2019a} and in two separate cavities~\cite{Yu2020a,Yu2020b}, and tripartite entanglement has also been studied by incorporating a mechanical oscillator~\cite{Li2018a}. However, two important elements are still missing in the investigations of magnon entanglement, i.e., {\em discrete variable} and {\em long distance}, which are directly relevant to various quantum information tasks.

In this Letter, we provide a systematic analysis of discrete-variable entanglement generation among magnon and superconducting qubits at the low excitation regime in two remote fiber-coupled hybrid cavity systems, see illustration in Fig.~\ref{fig_schm}. In each hybrid unit, a YIG sphere together with a superconducting qubit (SQ)~\cite{Tabuchi2016a,Tabuchi2015a,Lachance-Quirion2017a,sq_add}, reside in a microwave cavity. Surprisingly, there exists an optimal fiber distance that can achieve maximal magnon-magnon and magnon-SQ entanglements with the shortest amount of time. Experimentally realistic parameters and environmental noise effects are also considered to confirm the robustness of entanglement generation.

\section{Coupled hybrid magnonic system}

A magnon~\cite{Soykal2010a, Tabuchi2014a, White1983a} represents a quanta of spin waves hosted by a ferromagnetic material such as YIG, whose self-Hamiltonian is given by $g \mu_B B_z \hat{S}_z$, where $g$ is the electron $g$-factor, $\mu_B$ is the Bohr magneton, and $B_z$ is the effective magnetic field. The collective spin operator $\hat{S}_{x,y,z}$ may be expressed in terms of bosonic creation and annihilation operators, by means of a Holstein-Primakoff transformation~\cite{Holstein1940a}, $\hat{S}_+=m ^\dagger \left(\sqrt{2S-m ^\dagger m}\right)$, $\hat{S}_-=\left(\sqrt{2S-m ^\dagger m}\right) m$, and $\hat{S}_z=m ^\dagger m -S$. Here $m$ is a bosonic annihilation operator, $\hat{S}_{\pm}=\hat{S}_x \pm \hat{S}_y$, and $S$ is the total collective spin number. When $2S \gg \langle m ^\dagger m \rangle$, one has $\hat{S}_+ = \sqrt{2S} m ^\dagger$. The magnon can magnetically couple to a microwave cavity mode via Zeeman effect~\cite{Tabuchi2016a,Soykal2010a,Imamoglu2009a}, resulting in an interaction Hamiltonian $H_{\rm int}^{m \tdsh c} \propto \hat{S}_+ a + S_- a^\dagger$, where $a$ is the annihilation operator of the cavity mode, under the usual rotating wave approximation (RWA). Therefore, the effective coupling can be treated as exchange between the magnon and microwave photon, $H_{\rm int}^{m \tdsh c} = g_{\rm eff} (m^\dagger a + h.c.)$. Likewise, as an artificial atom~\cite{Xiang2013a}, a SQ may be fabricated to couple to the electric field of the cavity via the exchange of microwave photons, typically by placing the qubit where the electric field is close to the maximum~\cite{Tabuchi2016a}. While the magnon-cavity coupling is magnetic in nature, the SQ-cavity coupling is electric~\cite{You2003a,Tabuchi2016a,Koch2007a}. The canonical variables in a transmon SQ are the number operator $n$ and the phase difference $\varphi$, satisfying the commutation relationship $[\varphi,n]=i$, akin to the canonical position and momentum operators. The transom's dipole coupling to the cavity is proportional to the charge (the number operator) of the SQ, and the voltage of the cavity, $H_{\rm int}^{q \tdsh c} \propto n(a+a ^\dagger)$. By rewriting ${\varphi, n}$ in terms of bosonic creation and annihilation operators and taking the RWA, the coupling of the transmon to the cavity can be cast into a similar form as the magnon-cavity coupling.

As building blocks for a large scale network of coupled hybrid cavities, let us consider two microwave cavities connected via a communication channel. For long-distance information transfer, we propose to use an optical fiber accompanied by two reversible microwave-optical photon converters. Such a conversion can be realized by using nonlinear effects in opto-mechanical systems~\cite{conv_review} or doped rare-earth crystal in a cavity in conjunction with a suitably-tuned laser~\cite{conv_ion1,conv_ion2,conv_ion3}. In the latter case, a rare-earth doped crystal such as erbium-doped crystal may act as an effective three-level system, where the microwave excitation drives the transition from the ground state $|g \rangle$ to a first-excited state $|1\rangle$, and an external laser drives the transition from $|1 \rangle$ to the second excitation state $|2 \rangle$. Consequently, the transition from $|2 \rangle$ back to the ground state $|g \rangle$ gives an optical photon which is then transmitted over an optical fiber. The reverse steps are carried out at the other end to convert the optical photon back into a microwave one. Inside each of the microwave-cavity, there is a YIG sphere hosting the magnon and a SQ. The magnon and SQ are both coupled to the cavity mode, while the direct coupling between them is negligible~\cite{Tabuchi2016a,Tabuchi2015a}. The model is schematically shown in Fig.~\ref{fig_schm}, and the system Hamiltonian is given by

\begin{align}
    H_s &= \sum_{i=1,2} \left[\omega_c a_i ^\dagger a_i + \omega_m m_i^\dagger m_i + \omega_q b_i ^\dagger b_i  \right. \nonumber \\
    &\left. + g_m(m_i ^\dagger a_i + h.c.) + g_q(b_i ^\dagger a_i + h.c.) \right] + J(a_1^\dagger a_2+a_1 a_2 ^\dagger), \label{eq_hsys}
\end{align}
assuming identical cavities, where $a_i,\,m_i$ and $b_i$ are the annihilation operators for the cavity, magnon and SQ respectively, $\omega_{c(m,q)}$ is the frequency of the cavity (magnon, SQ), $g_{m(q)}$ is the coupling strength between the magnon (SQ) and the cavity, and $J$ is the overall effective coupling strength of the communication channel between the two cavities. When the channel between the two cavities is realized with frequency converters at both ends connected by an optical fiber (Fig.~\ref{fig_schm}), we have $J=\xi^2 J_{f}$, where $\xi$ is the conversion efficiency (for example, $\xi$ is demonstrated to be 100\% in the scheme proposed in Ref.~\cite{conv_ion1,conv_ion2,conv_ion3}) and $J_{f}$ is the coupling rate of the optical fiber. The coupling strength of the fiber may be dictated by the leakage or decay rate of the cavity mode $\Gamma_c$, estimated as~\cite{Pellizzari1997a} $J_f \approx \sqrt{8\pi c \Gamma_c/L}$, where $c$ is the speed of light and $L$ is the fiber's length. It is also noted that under specific conditions~\cite{Tabuchi2015a,Tabuchi2016a}, the cavity mode can be adiabatically eliminated to create an effective SQ-magnon coupling through exchanging virtual cavity photons. The above Hamiltonian will be analyzed exactly without specifying any parameter. Such a setup can accommodate a remote generation of entanglement between macroscopic objects, either being two magnons or being one SQ and one magnon (e.g. $q_1$ and $m_2$). Since the system Hamiltonian~\eqref{eq_hsys} conserves the total number of excitations, we will focus on the single-excitation manifold and study the discrete entanglement dynamics. This allows us to treat the magnons, SQs and cavity modes effectively as two-level systems (qubits), which can be potentially useful for various quantum information processing devices.

\begin{figure}
  \centering
      \includegraphics[scale=.40]{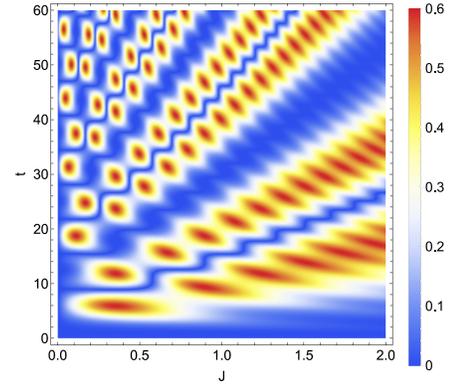}
      \caption{(Color online) Entanglement dynamics as measured by the concurrence between the two magnons in the resonant case, as a function of the fiber coupling strength $J$ and time $t$, with $g_m=0.4$ and $g_q=0.3$. It can be observed that there exists an optimal fiber coupling $J$ to reach peak entanglement in the shortest amount of time.}\label{fig_rmm}
  \end{figure} 

\section{Entanglement dynamics}

Now, we turn to the entanglement dynamics of the system. Without loss of generality, the initial state of the system is prepared as a separable state, with one excitation in the first SQ, $|\psi(0) \rangle=|001,000 \rangle$, where the basis of the system is $|c_1m_1q_1,c_2m_2q_2 \rangle$, where $c_{i}$ denotes cavity $i$ and the SQ (magnon) in cavity $i$ is denoted as $q_i(m_i)$, $i=1,2$. The system dynamics can be exactly solved, and the two-party reduced density operator of the combining SQs and the magnons takes an $X$-state~\cite{Yu2007a}, giving an analytical expression for the entanglement, measured by the concurrence~\cite{Wootters1998a}. In the case of a resonant configuration $\omega_c=\omega_q=\omega_m\equiv\omega$, the self-Hamiltonian of the system only gives an overall global phase in the single excitation manifold, thus the choice of $\omega$ does not affect the system dynamics. 

We plot the magnon-magnon entanglement dynamics for varying fiber coupling $J$ in Fig.~\ref{fig_rmm}. Surprisingly, stronger fiber coupling does not necessarily lead to faster entanglement generation. Indeed, there exists an optimal fiber coupling $J$ such that the time to generate peak entanglement is the shortest. It can be shown that the optimal time achieving peak entanglement is $t=2n\pi/\sqrt{G_0}$, where $G_0=4 \left(g_m^2+g_q^2\right)+J^2$ and $n$ is a positive integer, with a set of corresponding coupling strength $J$. Note that $J$ depends on $n$ and is also subject to other constraints (see supplemental document), so larger $J$ does not necessarily lead to faster entanglement generation. It can be shown that when $n=1$ the shortest time is given by $t_{\rm opt} = 2\pi/(3J_{\rm opt})$ with a corresponding optimal fiber coupling $J_{\rm opt}=\sqrt{(g_m^2+g_q^2)/2}$. The peak entanglement quantified by concurrence \cite{Wootters1998a} is obtained as
\begin{equation}
C_{m_1\tdsh m_2}|_{\rm opt}={3 \sqrt{3} r_q^2}/{2 \left(r_q^2+1\right)^2},
\end{equation}
which depends on the ratio of the SQ-cavity and magnon-cavity coupling strengths, i.e., $r_q=g_q/g_m$, see illustration by the red peaks in Fig.~\ref{fig_rmm}. For varying magnon(SQ)-cavity coupling rate $g_{m(q)}$, the peak concurrence is maximized when $g_m=g_q=J$, at $C_{m_1\tdsh m_2}|_{\rm max}=3\sqrt{3}/8$. This configuration corresponds to an equally-spaced (spacing being $J$) energy spectrum of the system Hamiltonian with a two-fold degeneracy at zero energy. A similar feature can also be observed for the SQ-SQ entanglement and the hybrid entanglement between the SQ and the remote magnon with shifted phases, representing a dynamical entanglement distribution \cite{Qian2005a, Qian2018a}. For the SQ-SQ entanglement, the peak entanglement is given by ${\sqrt{(\eta-1) (\eta+3)^3}}/{8 (r_q^2+1)^2}$, where $\eta=\sqrt{8 r_q^4+1}$. The peak SQ-SQ entanglement thus approaches to $1$ when $r_q \rightarrow +\infty$ due to the fact that the initial excitation is on $q_1$. It's also worth pointing out that the system under consideration~\eqref{eq_hsys} does not actually create Gaussian continuous-variable entanglement due to the lack of squeezing terms, but the exchange of excitations is enough to generate a discrete-variable entanglement.

\begin{figure}
  \centering
      \includegraphics[scale=.70]{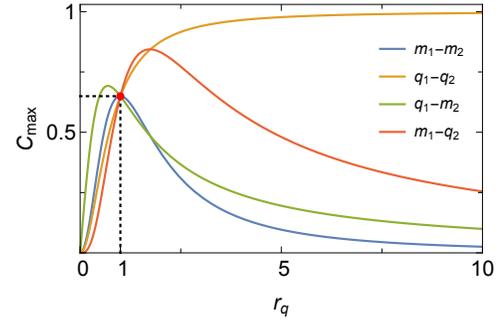}
      \caption{(Color online) Peak entanglement as a function of the SQ- and magnon-cavity coupling $r_q=g_q/g_m$. The SQ-SQ entanglement monotonically increases, and for others the maximum of the peak is reached at different $r_q$'s. It may also be observed that all curves coincide at $r_q=1$, $C=3\sqrt{3}/8$.}\label{fig_r_cx}
\end{figure}

For the hybrid entanglement between the remote SQ and magnons, although the system Hamiltonian is symmetric for SQ and magnon in each cavity, the choice of the initial state with one excitation at SQ $1$ breaks the symmetry, consequently, the entanglement of $q_1\tdsh m_2$ deviates from that of $m_1\tdsh q_2$. It can be readily shown that $C_{q_1\tdsh m_2} = C_{q_1\tdsh q_2}/r_q$ and $C_{m_1\tdsh q_2} = r_q C_{m_1\tdsh m_2}$. The peak concurrence is plotted in Fig.~\ref{fig_r_cx} as a function of $r_q$. Interestingly, we observe that, while the SQ-SQ concurrence asymptotically increases with $r_q$,  for the other types of entanglement there exists a maximum value for the peak concurrence at different coupling ratio $r_q$. All curves coincide at $r_q=1$, which is expected since when $g_m=g_q$, the system dynamics does not distinguish between the SQ and magnon. In the limit of $r_q \rightarrow \infty$, the SQ-SQ peak entanglement approaches $1$ while all others go to $0$. It should be noted that, while the time to generate peak concurrence is generically given by $t=2n\pi/\sqrt{G_0}$, the corresponding fiber coupling $J$ is different for different types of entanglements. For example, $C_{m_1\tdsh q_2}$ reaches its maximum $27/32$ at $r_q=\sqrt{3}$, while $C_{q_1 \tdsh m_2}$ reaches $\sim 0.6922$ at $r_q\approx 0.6896$.

\begin{figure}
    \centering
    \includegraphics[scale=.7]{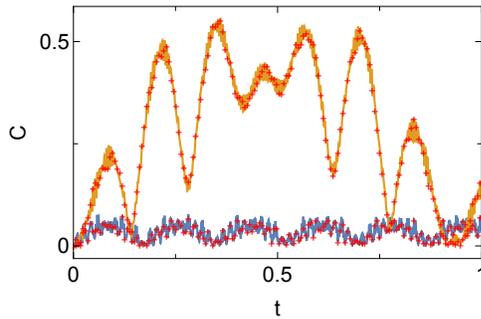}
    \caption{(Color online) Concurrence between the two magnons as a function of time, with $g_m/2\pi = 21 \mhz$, detuning $\delta/2\pi=183 \mhz$, SQ coupling $g_q/2\pi=117 \mhz$ (blue line) and $30 \mhz$ (orange line). The red crosses ($+$) indicate results obtained under a general non-Markovian open system dynamics with cavity leaking rate $\Gamma_c/2\pi=1.8\mhz$.}\label{fig_nr}
\end{figure}

Next, we consider the more general non-resonant situation. In light of physical implementation of our system, parameters that have been experimentally reported are used. We choose an experimentally accessible configuration~\cite{Tabuchi2016a} with the magnon-cavity coupling rate $g_m/2\pi = 21 \mhz$, the SQ-cavity coupling rate $g_q/2\pi=117 \mhz$, and let the frequencies of the SQ and magnon be equal. The detuning $\delta=\omega_c-\omega_q$ is set to be $183\times 2\pi \mhz$. With a typical cavity decay rate $\Gamma_c/2\pi=1.8\mhz$, we estimate~\cite{Pellizzari1997a} the fiber coupling strength $J \approx 92.3 \mhz$, where the length of the fiber is $10\;m$. 
In Fig.~\ref{fig_nr} we show that a quasi-periodic oscillation exists, and by tuning the SQ coupling rate to be closer to the magnon coupling rate at $g_q/2\pi=30 \mhz$, magnon entanglement generation can be greatly enhanced. 

Realistically, the system inevitably suffers dissipation and needs an open-system approach~\cite{Breuer2002a}. In this case, we model the environment of the cavities by two local multi-mode bosonic baths at zero temperature. The Hamiltonians of the bath and interaction are given by $H_{\rm bath} = \sum_{j=1,2} \sum_k \omega_k \envb{k}^\dagger \envb{k}$, and $H_{\rm int} = \sum_{j=1,2} \sum_k g_k L_j \envb{k}^\dagger + g_k^* L_j^\dagger \envb{k}$, where $H_{\rm int}$ describes the interaction between the cavity system and bath, $\omega_k$ is the frequency of the $k$-th bath mode, $g_k$ is the corresponding coupling strength, $L_j$ $(j=1,2)$ are the cavity-bath coupling operators and $\envb{k}$ denotes the $k$-th mode annihilation operator of bath $j$. For the leaky cavities with coupling operators $L_j=\Gamma_c a_j$, we numerically simulate the open system dynamics under generic colored noises by using a non-Markovian Schr\"{o}dinger equation~\cite{Yu1999a,Diosi1998a,Strunz1999a,Luo2015a}. The reduced density operator for the open system is recovered by averaging the quantum trajectories generated by the stochastic Schr\"{o}dinger equation with a typical bath memory function $\alpha(t,s)=\sum_k |g_k|^2 e^{-i \omega_k (t-s)}=\exp(-\gamma |t-s|)/2$. The time dependent entanglement is shown to be robust against environmental noises as displayed in Fig.~\ref{fig_nr} with $\gamma=0.7$ and $\Gamma_c/2\pi=1.8 \mhz$. When the cavity dissipation becomes larger, one may employ an external quantum control scheme \cite{Jing2015a} to mitigate the effect.

\section{Discussion and conclusion}
We study remote entanglement generation of macroscopic magnons and superconducting qubits in two coupled hybrid microwave cavity systems. The communication channel of the two cavities is proposed to be achieved with two microwave-optical photon converters and a conventional optical fiber. It is shown that entanglement can be reliably created between the remote parties. Remarkably, there exists an optimal fiber coupling strength to achieve peak entanglement in the least amount of time, and stronger fiber couplings do not necessarily mean a faster entanglement generation. Therefore, to accommodate long-distance entanglement generation, one does not need extra strong coupling, but rather the optimal coupling strength as a function of the magnon and qubit-cavity coupling rate. Moreover, the effect of non-Markovian environment is studied. The system is shown to be robust against the noise.

By encoding information with discrete variables rather than continuous variables, this hybrid system has provided a new framework to exploit the compatibility of SQs with a microwave cavity and the longevity, operability and potential scalability of magnon SQs. Since quantum entanglement is not a direct physical observable, quantum state tomography~\cite{tomo1} may be applied on the qubit or magnon to measure and detect the entanglement generation in our system. It would be interesting to study multipartite entanglement as well as thermal bath effects in the proposed coupled hybrid systems~\cite{ref_thermal}.



\section*{Funding}

This work is supported by NSF PHY-0925174.

\section*{Disclosures}

The authors declare no conflicts of interest.




\newpage 

\renewcommand\refname{Full References}



\section*{Supplementary material}

# Nonlocal magnon entanglement generation in coupled hybrid cavity systems: supplemental document

In this supplemental document, we provide detailed calculations of entanglement dynamics, optimal fiber coupling strength, and the treatment of non-Markovian open system dynamics.

## 1. STATE AND ENTANGLEMENT DYNAMICS IN THE CLOSED SYSTEM CASE

For the unitary evolution under the Hamiltonian Eq. (1) in the main text, first note that the Hamiltonian conserves the total number of excitations, which enables us to focus on the single-excitation subspace. In the single-excitation subspace $\{|100,000\rangle, |010,000\rangle \ldots |000,001\rangle\}$, where the basis of the system is $|c_1 m_1 q_1, c_2 m_2 q_2\rangle$ with $c_i$ denotes cavity $i = 1, 2$ and the SQ (magnon) in cavity $i$ is denoted as $q_i(m_i)$. The Hamiltonian can be expressed in this basis as

$$H = \begin{pmatrix} \omega_c & g_m & g_q & J & 0 & 0 \\ g_m & \omega_m & 0 & 0 & 0 & 0 \\ g_q & 0 & \omega_q & 0 & 0 & 0 \\ J & 0 & 0 & \omega_c & g_m & g_q \\ 0 & 0 & 0 & g_m & \omega_m & 0 \\ 0 & 0 & 0 & g_q & 0 & \omega_q \end{pmatrix}. \tag{S1}$$

In the resonant case $\omega_c = \omega_m = \omega_q = \omega$, $\omega$ is a trivial diagonal constant and can therefore be dropped. The initial state is taken to have one excitation in the qubit 1, i.e., $|\psi(0)\rangle = |001, 000\rangle$. The state at time $t$ is then given by

$$|\psi(t)\rangle = \begin{pmatrix} -\frac{2i g_q \cos\left(\frac{Jt}{2}\right) \sin\left(\frac{\Omega t}{2}\right)}{\Omega} \\ \frac{g_m g_q \left(J \sin\left(\frac{Jt}{2}\right) \sin\left(\frac{\Omega t}{2}\right) + \Omega \cos\left(\frac{Jt}{2}\right) \cos\left(\frac{\Omega t}{2}\right) - \Omega\right)}{\Omega(g_m^2 + g_q^2)} \\ \frac{g_q^2 \left(\frac{J \sin\left(\frac{Jt}{2}\right) \sin\left(\frac{\Omega t}{2}\right)}{\Omega} + \cos\left(\frac{Jt}{2}\right) \cos\left(\frac{\Omega t}{2}\right)\right) + g_m^2}{g_m^2 + g_q^2} \\ -\frac{2 g_q \sin\left(\frac{Jt}{2}\right) \sin\left(\frac{\Omega t}{2}\right)}{\Omega} \\ -\frac{i g_m g_q \left(\sin\left(\frac{Jt}{2}\right) \cos\left(\frac{\Omega t}{2}\right) - \frac{J \cos\left(\frac{Jt}{2}\right) \sin\left(\frac{\Omega t}{2}\right)}{\Omega}\right)}{g_m^2 + g_q^2} \\ -\frac{i g_q^2 \left(\sin\left(\frac{Jt}{2}\right) \cos\left(\frac{\Omega t}{2}\right) - \frac{J \cos\left(\frac{Jt}{2}\right) \sin\left(\frac{\Omega t}{2}\right)}{\Omega}\right)}{g_m^2 + g_q^2} \end{pmatrix}, \tag{S2}$$

where $\Omega = \sqrt{4\left(g_m^2 + g_q^2\right) + J^2}$.

The concurrence between the magnons (qubits/cavities/qubit1-magnon2) ($C_m, C_q, C_c, C_{qm}$) can be easily obtained. The reduced density operator for the two magnons (in computational basis

$|11\rangle, |10\rangle, |01\rangle, |00\rangle)$ is given as

$$\rho_{mm} = \begin{pmatrix} 0 & 0 & 0 & 0 \\ 0 & x & w & 0 \\ 0 & w^* & y & 0 \\ 0 & 0 & 0 & z \end{pmatrix}, \quad \text{(S3)}$$

where

$$x = \frac{g_m^2 g_q^2 \left[ J \sin\left(\frac{Jt}{2}\right) \sin\left(\frac{\Omega t}{2}\right) + \Omega \cos\left(\frac{Jt}{2}\right) \cos\left(\frac{\Omega t}{2}\right) - \Omega \right]^2}{\Omega^2 \left(g_m^2 + g_q^2\right)^2}, \quad \text{(S4)}$$

$$y = \frac{g_m^2 g_q^2 \left[ \Omega \sin\left(\frac{Jt}{2}\right) \cos\left(\frac{\Omega t}{2}\right) - J \cos\left(\frac{Jt}{2}\right) \sin\left(\frac{\Omega t}{2}\right) \right]^2}{\Omega^2 \left(g_m^2 + g_q^2\right)^2}, \quad \text{(S5)}$$

$$z = 1 - x - y, \quad \text{(S6)}$$

$$w = \frac{i g_m^2 g_q^2 \left[ \Omega \sin\left(\frac{Jt}{2}\right) \cos\left(\frac{\Omega t}{2}\right) - J \cos\left(\frac{Jt}{2}\right) \sin\left(\frac{\Omega t}{2}\right) \right]}{\Omega^2 \left(g_m^2 + g_q^2\right)^2}$$

$$\times \left[ J \sin\left(\frac{Jt}{2}\right) \sin\left(\frac{\Omega t}{2}\right) + \Omega \cos\left(\frac{Jt}{2}\right) \cos\left(\frac{\Omega t}{2}\right) - \Omega \right]. \quad \text{(S7)}$$

This is a mixed state in X form. Its concurrence can be obtained directly as

$$C_m = 2 \frac{g_m^2 g_q^2}{G_0 \left(g_m^2 + g_q^2\right)^2} \left| \left[ \sqrt{G_0} \cos\frac{\sqrt{G_0}t}{2} \sin\frac{Jt}{2} - J \cos\frac{Jt}{2} \sin\frac{\sqrt{G_0}t}{2} \right] \right|$$

$$\times \left| \left[ J \sin\frac{Jt}{2} \sin\frac{\sqrt{G_0}t}{2} + \sqrt{G_0} \left( \cos\frac{\sqrt{G_0}t}{2} \cos\frac{Jt}{2} - 1 \right) \right] \right|, \quad \text{(S8)}$$

where $G_0 = \Omega^2 = 4\left(g_m^2 + g_q^2\right) + J^2$.

## 2. OPTIMAL TIME AND COUPLING FOR REACHING PEAK ENTANGLEMENT

In this section we provide the details of achieving peak entanglement with optimal time and fiber coupling through the analysis of the time dependent entanglement given as $C_m$ in Eq. (S8). We denote $C_m = |\widetilde{C}_m|$, and proceed to find the extrema of $\widetilde{C}_m$ through the partial derivative

$$\partial_t \widetilde{C}_m \propto \sin\frac{\sqrt{G_0}t}{2} \left[ J \sin\left(\frac{\sqrt{G_0}t}{2}\right) \cos(Jt) + \sqrt{G_0} \left( \sin\left(\frac{Jt}{2}\right) - \cos\left(\frac{\sqrt{G_0}t}{2}\right) \sin(Jt) \right) \right]. \quad \text{(S9)}$$

The global minimum/maximum may occur along $t$ where $\sin\frac{\sqrt{G_0}t}{2} = 0$. The maximum can be reached when $t = 2n\pi/\sqrt{G_0}$, subject to an optimal $J$. To find this optimal coupling value, we observe that at such times, the concurrence is given by

$$C_m|_t = 2 \left| \frac{g_m^2 g_q^2 \left(1 - (-1)^n \cos\frac{Jn\pi}{\sqrt{G_0}}\right) \sin\frac{Jn\pi}{\sqrt{G_0}}}{\left(g_m^2 + g_q^2\right)^2} \right| \quad \text{(S10)}$$

$$= \frac{2 g_m^2 g_q^2}{\left(g_m^2 + g_q^2\right)^2} \left| \left[1 - (-1)^n \cos x\right] \sin x \right|, \quad \text{(S11)}$$

where $x = \frac{Jn\pi}{\sqrt{G_0}}$. For odd $n$, the maximum of $|(1 + \cos x) \sin x| = 3\sqrt{3}/4$ is reached when $\cos x = 1/2$; for even $n$, the maximum of $|(1 - \cos x) \sin x| = 3\sqrt{3}/4$ is reached when $\cos x = -1/2$. We'll

get the same result if we calculate $\partial_J [C_m|_t] = 0$, which gives $\cos(x) = -(-1)^n/2$ (corresponds to max) and $\cos(x) = (-1)^n$ (corresponds to $C_m = 0$). Therefore, the optimal $J$ is given by:

$$J = 2(6m+1)\sqrt{\frac{g_m^2 + g_q^2}{9n^2 - (6m+1)^2}}, \quad 6m+1 < 3n, \text{odd } n \tag{S12}$$

$$J = 2(6m+5)\sqrt{\frac{g_m^2 + g_q^2}{9n^2 - (6m+5)^2}}, \quad 2m+2 \leq n, \text{odd } n \tag{S13}$$

or

$$J = 4(3m+1)\sqrt{\frac{g_m^2 + g_q^2}{9n^2 - 4(3m+1)^2}}, \quad 2m+1 \leq n, \text{even } n \tag{S14}$$

$$J = 4(3m+2)\sqrt{\frac{g_m^2 + g_q^2}{9n^2 - 4(3m+2)^2}}, \quad 6m+4 < 3n, \text{even } n \tag{S15}$$

where $m$ is a non-negative integer. Labeling the cases of Eq. (S12) through Eq. (S15) as $c1\ldots c4$, the corresponding optimal time $t = 2n\pi/\sqrt{G_0}$, we have the optimal times given by

$$t_{c1} = \frac{1}{3}\pi\sqrt{\frac{9n^2 - (6m+1)^2}{g_m^2 + g_q^2}}, \quad 6m+1 < 3n, \text{odd } n \tag{S16}$$

$$t_{c2} = \frac{1}{3}\pi\sqrt{\frac{9n^2 - (6m+5)^2}{g_m^2 + g_q^2}}, \quad 2m+2 \leq n, \text{odd } n \tag{S17}$$

$$t_{c3} = \frac{1}{3}\pi\sqrt{\frac{9n^2 - 4(3m+1)^2}{g_m^2 + g_q^2}} \quad 2m+1 \leq n, \text{even } n \tag{S18}$$

$$t_{c4} = \frac{1}{3}\pi\sqrt{\frac{9n^2 - 4(3m+2)^2}{g_m^2 + g_q^2}} \quad 6m+4 < 3n, \text{even } n \tag{S19}$$

It can then be readily shown, for case 1, we write odd $n = 2*u + 1$, where $u$ is a non-negative integer,

$$t_{c1} = \frac{1}{3}\pi\sqrt{\frac{9(2u+1)^2 - (6m+1)^2}{g_m^2 + g_q^2}}, \quad 3m < 3u+1. \tag{S20}$$

Since $u > m - 1/3$ and $u$ is an integer, we have $u \geq m$, so

$$t_{c1} \geq \frac{1}{3}\pi\sqrt{\frac{9(2m+1)^2 - (6m+1)^2}{g_m^2 + g_q^2}}$$

$$\geq \frac{2}{3}\pi\sqrt{\frac{6m+2}{g_m^2 + g_q^2}}$$

$$\geq \frac{2\sqrt{2}\pi}{3\sqrt{g_m^2 + g_q^2}} \tag{S21}$$

For case 2, since $n$ is odd, one has $n \geq 2m+2 \Rightarrow n \geq 2m+3$. Combining with the fact that $2m+2$ is even, one can obtain

$$t_{c2} \geq \frac{1}{3}\pi\sqrt{\frac{9(2m+3)^2 - (6m+5)^2}{g_m^2 + g_q^2}}$$

$$\geq \frac{2}{3}\sqrt{2}\pi\sqrt{\frac{6m+7}{g_m^2 + g_q^2}}$$

$$\geq \frac{2\sqrt{14}\pi}{3\sqrt{g_m^2 + g_q^2}}. \tag{S22}$$

3

Likewise, for case 3, since $n$ is even, one has $n \geq 2m + 1 \Rightarrow n \geq 2m + 2$. Combining with the fact that $2m + 1$ is odd, one can achieve

$$t_{c3} \geq \frac{1}{3}\pi \sqrt{\frac{9(2m+2)^2 - 4(3m+1)^2}{g_m^2 + g_q^2}}$$
$$\geq \frac{4}{3}\pi \sqrt{\frac{3m+2}{g_m^2 + g_q^2}}$$
$$\geq \frac{4\sqrt{2}\pi}{3\sqrt{g_m^2 + g_q^2}}. \tag{S23}$$

For case 4, since $n$ is even, we write $n = 2u$, where $u$ is a positive integer, thus

$$t_{c4} = \frac{2}{3}\pi \sqrt{\frac{9u^2 - (3m+2)^2}{g_m^2 + g_q^2}}, \quad 3m + 2 < 3u. \tag{S24}$$

Since $u > m + 2/3$ and $u$ is an integer, we have $u \geq m + 1$,

$$t_{c4} \geq \frac{2}{3}\pi \sqrt{\frac{9(m+1)^2 - (3m+2)^2}{g_m^2 + g_q^2}}$$
$$\geq \frac{2}{3}\pi \sqrt{\frac{6m+5}{g_m^2 + g_q^2}}$$
$$\geq \frac{2\sqrt{5}\pi}{3\sqrt{g_m^2 + g_q^2}} \tag{S25}$$

Thus, the *global* minimum time is reached for case 1 when $n = 1$, $m = 0$, and is given by

$$t = \frac{2\sqrt{2}\pi}{3\sqrt{g_m^2 + g_q^2}}, \tag{S26}$$

with a corresponding optimal coupling

$$J = \frac{\sqrt{g_m^2 + g_q^2}}{\sqrt{2}}, \tag{S27}$$

and the maximum concurrence is

$$C_m|_{\max} = \frac{3\sqrt{3}g_m^2 g_q^2}{2\left(g_m^2 + g_q^2\right)^2}. \tag{S28}$$

Along with the result $J_f \approx \sqrt{8\pi c \Gamma_c / L}$ in Ref. [1], these expressions above allow us to estimate the length of the fiber for optimal entanglement generation. For example, with couplings $g_m/2\pi = 21$ MHz, $g_q/2\pi = 30$ MHz, and leaking rate $\Gamma_c/2\pi = 1.8$ MHz, the distance can be approximately 32 meters, generating a peak concurrence of around 0.57 in about 41 nanoseconds (corresponding to $n = 3$, $m = 0$). Under the same parameters, if one is willing to wait about 259 nanoseconds (corresponding to $n = 19$, $m = 0$), the corresponding distance is about 1307 meters, with the same peak concurrence.

For varying $g_{m(q)}$, $C_m|_{\max}$ has a maximum value of $\frac{3\sqrt{3}}{8}$, reached when $|g_m| = |g_q|$. It's noteworthy that in this case, $g_m = g_q = J$, and the eigen-energy of the system Hamiltonian become evenly-spaced (with a two-fold degeneracy at zero energy)

$$E_{1\ldots 6} = \{-2J, -J, 0, 0, J, 2J\}. \tag{S29}$$

Similar features can be found for qubit-qubit and qubit-magnon entanglement.

4

## 3. OPEN SYSTEM DYNAMICS

For the non-Markovian open system dynamics, we consider the system embedded in an infinite-mode bosonic bath, where the total Hamiltonian is given by

$$H = H_s + H_{\text{bath}} + H_{\text{int}}, \tag{S30}$$

where $H_{\text{bath}} = \sum_{j=1,2} \sum_k \omega_k \bar{b}_{j,k}^\dagger \bar{b}_{j,k}$, $H_{\text{int}} = \sum_{j=1,2} \sum_k g_k L_j \bar{b}_{j,k}^\dagger + g_k^* L_j^\dagger \bar{b}_{j,k}$. $H_{\text{int}}$ describes the interaction between the cavity system and bath, $\omega_k$ is the frequency of the $k$-th bath mode, $g_k$ is the corresponding coupling strength, $L_j$ ($j = 1, 2$) are the cavity-bath coupling operators and $\bar{b}_{j,k}$ denotes the $k$-th mode annihilation operator of bath $j$. Here, we consider leaky cavities with the coupling operators $L_j = \Gamma_c a_j$. The quantum state diffusion (QSD) equation which describes a set of stochastic trajectories [2] is used to calculate the non-Markovian dynamics. Projecting the two baths onto the Bargmann coherent state $|z^{(j)}\rangle = e^{z_{k,j} \bar{b}_{j,k}}|0\rangle$, $j = 1, 2$, we have the QSD equation

$$\partial_t |\psi_t(z_1, z_2)\rangle = \left[ -iH + L_1 z_1^*(t) + L_2 z_2^*(t) - L_1^\dagger \bar{O}_1(t) - L_2^\dagger \bar{O}_2(t) \right] |\psi_t(z_1, z_2)\rangle, \tag{S31}$$

where $z_j^*(t) \equiv -i \sum_k g_k z_{k,j}^* e^{i\omega_k t}$, and the leading order noise-free $O_{1,2}(t, s)$ satisfies

$$\partial_t O_i(t, s) = \left[ -iH - L_1^\dagger \bar{O}_1(t) - L_2^\dagger \bar{O}_2(t), O_i(t, s) \right], \tag{S32}$$

where $\bar{O}_i(t) = \int_0^t \alpha(t, s) O_i(t, s) ds$, $i = 1, 2$. The reduced density operator for the system is given by an ensemble average of the trajectories, $\rho_s(t) = \mathcal{M}[|\psi_t(z_1, z_2)\rangle \langle \psi_t(z_1, z_2)|]$. This leads to the non-Markovian master equation

$$\partial_t \rho_s(t) = [-iH, \rho_s(t)] + \left[ L_1, \rho_s(t) \bar{O}_1^\dagger \right] + \left[ L_2, \rho_s(t) \bar{O}_2^\dagger \right] - \left[ L_1^\dagger, \bar{O}_1 \rho_s(t) \right] - \left[ L_2^\dagger, \bar{O}_2 \rho_s(t) \right]. \tag{S33}$$




\end{document}